\documentclass[12pt]{article}

\usepackage{graphicx}
\usepackage{dcolumn}
\usepackage{bm}
\usepackage{amssymb}
\usepackage{enumerate}

\usepackage{graphicx} 
\begin{document}

\title{On the ultraviolet behavior of the invariant charge  in
quantum electrodynamics} 
\author{N.V.Krasnikov
\\INR RAS,   Moscow 117312, Russia
\\and
\\ JINR, Dubna 141980, Russia
}
\maketitle
\begin{abstract}
  In this paper we study  the ultraviolet behavior of the invariant charge  in QED.
  We show that for complex momenta the invariant charge does not have Landau pole singularity. We can define new invariant charge
  as real part of standard invariant charge.
  New invariant charge is limited  from above and does not have Landau pole singularity. Also we use the $1/N$ perturbation
  theory  for the investigation of the ultraviolet behavior of the  invariant charge. To this aim we consider QED with imaginary charge
  which is asymptotically free but nonphysical model. In QED with nonphysical imaginary charge we can reliably calculate the
       ultraviolet asymptotics for the    $(1/N)^k$ correction to the invariant charge, namely:  $\alpha_k(\frac{p^2}{\mu^2}, \alpha) \sim (\ln(\frac{p^2}{\mu^2}))^{-k-1}$
  at $k > 1$ and $\alpha_1(\frac{p^2}{\mu^2}, \alpha) \sim (\frac{\ln(\ln(\frac{p^2}{\mu^2})}{\ln^2(\frac{p^2}{\mu^2})})$  at $k =1$. The $1/N$ perturbation theory
  coincides  for QED with imaginary charge and standard QED with real charge.
  It means in particular that ultraviolet behavior of  the  $(1/N)^k$ correction  $\alpha_k(\frac{p^2}{\mu^2}, \alpha)$
   in real QED
coincides with the corresponding asymptotics for QED with imaginary charge.  We
  propose also to use the modified $1/N$ expansion which is ultraviolet finite. The comparison of the
  standard QED and nonphysical QED with imaginary charge gives hint that in other non asymptotically free models like
  supersymmetric QED, scalar QED or Wess-Zumino model ultraviolet asymptotics of the invariant charge coincides with
  leading log approximtion.

\end{abstract}
\newpage
\section{Introduction}

In quantum field theory
renormalizable models
with single coupling connstant  
are divided into asymptotically free and asympltotically non free models. Two famous examples  of asymptotically free and
asymptotically non free models are QCD (quantum chromodynamics) and  QED (quantum electrodynamics). In QCD  the effective
coupling constant decreases at small distances, therefore  we can use the perturbation theory in the ultraviolet region. The theory looks
self-consistent at least at small distances. While in QED the effective coupling constant increases in the  ultraviolet region 
and the perturbation theory is not applicable. Moreover the ``naive'' use of the perturbation theory leads to the
appearance of the famous Landau pole singularity.
There is  some  evidence \cite{EVIDENCE}
  but not a proof \cite{bog1}   that QED is
  not self-consistent local quantum field theory at small distances.

  In this paper we study  the ultraviolet behavior of the invariant charge  in QED. In QED the invariant charge is proportional to
  the photon propagator.  We show that for complex momenta the invariant charge does not have Landau pole singularity
  at leat in finite loops approximation. We point out that we can define new invariant charge as real part of standard invariant charge.
  New invariant charge is limited  from above and does not have Landau pole singularity. Also we use the $1/N$ perturbation
  theory \cite{N1, N2, N3}  for the investigation of the ultraviolet behaviour of the  invariant charge. To this aim we consider QED with imaginary charge
  which is asymptotically free but nonphysical model. In QED with imaginary charge we can reliably calculate the
  ultraviolet asymptotics of the $(1/N)^k$ correction to the invariant charge which is
  equal to $\alpha_k(\frac{p^2}{\mu^2}, \alpha) \sim (\ln(\frac{p^2}{\mu^2})^{-k-1}$
  at $k > 1$ and $\alpha_{k=1}(\frac{p^2}{\mu^2}, \alpha) \sim (\frac{\ln(\ln(\frac{p^2}{\mu^2})}{\ln^2(\frac{p^2}{\mu^2})})$  at $k = 1$.
    The $1/N$ perturbation theory
  coincides
     for QED with imaginary charge and standard QED with real charge.
  It means in particular that ultraviolet behavior of
  the invariant charge in standard   QED $\bar{\alpha}(\frac{p^2}{\mu^2}, \alpha)  \sim (\ln(\frac{p^2}{\mu^2}))^{-1} $
  is determined by   the leading log approximation. We
  propose also to use the modified $1/N$ expansion which is ultraviolet finite. The comparison of the
  standard QED and nonphysical QED with imaginary charge gives hint that in other non asymptotically free models like
  supersymmetric QED,  scalar QED  or Wess-Zumino model ultraviolet asymptotics of the invariant charge coincides with
  leading log approximtion.

  The organization of the paper is the following.  In the next section
  we consider the invariant charge in QED at complex momenta.
    In the  section 4  we
  investigate the QED ultraviolet behavior using the $1/N$ expansion.
    Section 5 contains concluding remarks.
  Appendixes A and B consider the N expansion in QED and QED
  with imaginary charge correspondingly.

  \section{The invariant charge at complex momenta}

  In this paper we  study   QED with N identical massless  fermions  $ \psi_k $. The interaction   Lagrangian
   has the form
     \begin{equation}
       L_{int} = \frac{e}{\sqrt{N}} \sum_{k=1}^{N}\bar{\psi}_k\gamma^{\mu} \psi_k A_{\mu} \,.
         \label{Nqed1}
     \end{equation}
  It is well known \cite{bog1} that the invariant charge in QED can be chosen  proportional to the transverse part of photon
  propagator
  with the normalization
   condition $  \bar{\alpha}(x= 1, \alpha, N) = \alpha $. Here $x = \frac{p^2}{\mu^2}$ and $\alpha = \frac{e^2}{4\pi}$.
   We consider QED in euclidean space-time. The invariant charge can be represented in the form
   \begin{equation}
     \bar{\alpha}(x, \alpha, N) =  \frac{\alpha}{1 + \alpha \Pi(x,\alpha, N)} \,
     \label{QEDa}
   \end{equation}
   In perturbation theory the  KL (Kallen-Lehmann) representation \cite{KL1,KL2}
   for  $\Pi(x,\alpha, N) $
    with one subtraction  \cite{Berestetsky}
   \begin{equation}
     \Pi(x,\alpha, N) = \int^{\infty}_{0} [\frac{\rho_{\Pi}(t, \alpha, N)}{t + x}   - \frac{\rho_{\Pi}(t, \alpha, N)}{t + 1}] dt      \,
     \label{QEDb}
     \end{equation}
   is valid.
     The renormalization group equation for the invariant charge  $\bar{\alpha}(x, \alpha, N)   $ has the form
   \begin{equation}
  x \frac{d}{dx}\bar{\alpha}(x, \alpha, N) = \psi(\bar{\alpha}, N) \,,
   \label{QEDb1}
   \end{equation}
   where the GLM (Gell-Mann ~-~Low) function  $\psi(\bar{\alpha}, N)$    \cite{Gel, Pet}    can be represented in the form
   \footnote{There are rigorous inequalities for the GLM function in QED \cite{Kr1} - \cite{Kr3} based on the use of the KL representation
     which are violated in QED with $N \gg 1 $ flavors.}
   \begin{equation}
 \psi(\bar{\alpha}, N) = \beta_2\alpha^2 +\sum_{k=1}^{\infty}(1/N)^k\psi_k(\alpha) \,.
     \label{QREDb2}
   \end{equation}
    Here      $\beta_2 = \frac{1}{3\pi}$   and
      $ \psi_k(\alpha) = \sum_{l =k+2}^{\infty} c_{l,k} \alpha^l  $.
   Note that in refs.\cite{QEDN1}-\cite{QEDN4} the first $1/N$ correction to the GLM function
   $ \psi_1(\alpha)$    have been calculated in
   different renormalization schemes\footnote{In ref.\cite{LINDE} authors used $1/N$ expansion and argued that in QED the vacuum is unstable as
    a  manifestation of the Landau pole singularity.}. In refs.\cite{Kat1,Kat2,Kat3} the GLM function has been calculated in four and five loop
   approximation for  different renormalization schemes.
 For QED with massless fermions the invariant charge $\bar{\alpha}(\frac{p^2}{\mu^2}, \alpha, N) $ is a function of
    the variables $\frac{p^2}{\Lambda^2}$ and N.
Here  the scale $\Lambda^2  = \mu^2\exp(-\int\frac{d\alpha}{\psi(\alpha, N)}) $ obeys
the equation $(\mu^{2}\frac{d}{d\mu^2} + \psi(\alpha, N)\frac{d}{d\alpha})\Lambda^2 =0 $. As a consequence  the invariant charge at
complex $p^2 = \exp(i\phi)|p^2|$ is equal to the invariant charge at real $|p^2|$ and complex $\Lambda^2 = |\Lambda|^2 \exp(-i\phi)$.   

In this section we consider the invariant charge  at complex $p^2$. Note that in refs.\cite{PIVO,PIVO1} the effects of the analytical continuation to complex momenta
have been studied in QCD.
  The main difference between euclidean region with  $Im~ p^2 = 0, ~p^2 \geq 0$ and the complex region is that in complex region the
  effective coupling constant does not have Landau pole singularity at least in finite loop approximation.
  Really, in one-loop approximation the
effective coupling constant obeys the equation
\begin{equation}
  \frac{d\bar{\alpha}}{dt} = \beta_2 \bar{\alpha}^2   \,,
    \label{C1}
    \end{equation}
    where $t = \ln(\frac{p^2}{\mu^2})$. We  use the normalization condition  $\bar{\alpha}(t = 0, \alpha) = \alpha$   and consider the case  $\beta_2  > 0$.
    The solution of the equation (\ref{C1}) has the well known form
    \begin{equation}
      \bar{\alpha}(\frac{p^2}{\mu^2}, \alpha) = \frac{\alpha}{1 - \alpha \beta_2  \ln (\frac{p^2}{\mu^2}) } =
        \frac{1}{-\beta_2\ln(\frac {p^2}{\Lambda^2})} \,,
        \label{C2}
    \end{equation}
    where $\Lambda^2 = \exp(\frac{1}{\alpha \beta_2})\mu^2$. The invariant charge  (\ref{C2}) is infinite at $p^2 = \Lambda^2 $
    (the famous Landau pole singularity), positive at $ p^2 < \Lambda^2 $ and negative at $p^2  > \Lambda^2 $. Consider the invariant charge (\ref{C2})
    at complex $p^2 = \exp(i \phi)|p^2|$.  Here 
      $ 0 < \phi \leq \pi $. For complex $p^2$ the invariant charge  (\ref{C2}) has the form
    \begin{equation}
      \bar{\alpha}(\frac{\exp(i\phi)|p^2|}{\Lambda^2}) = 
        \frac{\ln(\frac{|p^2|}{\Lambda^2}) -i\phi)}    
          {-\beta_2 ( \ln^2(\frac{|p^2|}{\Lambda^2}) + \phi^2)} \,.
        \label{C3}
    \end{equation}
    The invariant charge  (\ref{C3})  does not have Landau pole singularity  at complex $p^2$ including negative $p^2$.
    The absence of Landau pole singularity at complex $p^2$  is also valid in many loop approximation for beta function
    $\beta(\alpha) = \sum_{k=2}^{N}\beta_{k}\alpha^k$. For instance, in two-loop approximation the solution of the
    renormalization group equation can be  presented in the form
    \begin{equation}
      \bar{\alpha}(\frac{p^2}{\Lambda^2}) = \frac{1}{-\beta_2\ln(\frac {p^2}{\Lambda^2})}(1 +  \frac{\beta_3}{\beta^2_2}
      \frac{\ln(-\ln(\frac {p^2}{\Lambda^2}))}{\ln(\frac {p^2}{\Lambda^2})})  +
      O( \frac{1}{\ln^2(\frac {p^2}{\Lambda^2})}     )  \,.
      \label{C4}
    \end{equation}
       The invariant charge  (\ref{C4})  does not have Landau pole singularity at complex $p^2$.
     In one-loop approximation the real part of the invariant charge  (\ref{C3}) is
     \begin{equation}
         \bar{\alpha}_{Re, \phi}(\frac{|p^2|}{\Lambda^2}) \equiv Re~  \bar{\alpha}(\frac{\exp(i\phi)|p^2|}{\Lambda^2}) = 
        \frac{\ln(\frac{|p^2|}{\Lambda^2})}    
          {-\beta_2 ( \ln^2(\frac{|p^2|}{\Lambda^2}) + \phi^2)} \,.
        \label{C5}
     \end{equation}
     The most interesting case $\phi = \pi$ corresponds to time-like region $p^2 < 0$. At time-like region the real part of the invariant charge
     has maximum(minimum) at $\ln(\frac{|p^2|}{\Lambda^2}) = -\pi(\pi)$ equal to $\frac{1}{2\pi\beta_2}( -\frac{1}{2
       \pi\beta_2})$.
     We can define the invariant charge in time-like region as real part of the invariant charge 
       , i.e.   $\bar{\alpha}_{Re, \phi} = \frac{\bar{\alpha} + \bar{\alpha^*}}{2}$.
In  two-loop approximation (\ref{C4}) for $\phi = \pi$ and 
     $\ln(\frac{|p^2|}{\Lambda^2}) =   - \pi$
\begin{equation}
  \bar{\alpha}_{Re, \phi = \pi}(\frac{|p^2|}{\Lambda^2} = \exp(-\pi)) = \frac{1}{2\pi \beta_2}
    - \frac{\beta_3}{8\pi\beta_2^3}  \,.
   \label{C5a}
\end{equation}
In QED with the Lagrangian (\ref{Nqed1})   $\beta_2 = \frac{1}{3\pi}$,  $\beta_3 = \frac{1}{4N\pi^2}$ and two-loop correction    at (\ref{C5a})  is
less than 10 percent for $N  \geq 6$. It means that we can trust one-loop approximation 
   for $\bar{\alpha}_{Re, \phi}(\frac{|p^2|}{\lambda^2}, N)$  at $N \gg 1 $.

It is interesting to note that at  imaginary $p^2 = ix, ~Im ~x = 0$       the KL representation  for
$\bar{\alpha}_{Re, \phi= \frac{\pi}{2}}(\frac{|p^2|}{\Lambda^2})$ can be represented in the form
\begin{equation}
\bar{\alpha}_{Re, \phi = \frac{\pi}{2}}(     \frac{x}{\Lambda^2}) = x^2\int_{0}^{\infty}\frac{\rho(t)dt}{x^2 + t^2} \,,
  \label{C7}
\end{equation}
where $\rho(t) \geq 0$.
As a consequence of the representation (\ref{C7})  $\bar{\alpha}_{Re, \phi = \frac{\pi}{2}}(x)$ is increasing
function at variable  $x$, i.e. $  \frac{d\bar{\alpha}_{Re, \phi = \frac{\pi}{2}}(\frac{x}{\Lambda^2})}{dx} \geq 0 $. It means  that
one-loop approximation (\ref{C5}) is not applicable at $\frac{|p^2|}{\Lambda^2} \geq exp(-\frac{\pi}{2})$.

So we have found that the invariant charge at complex $p^2$ including negative $p^2$ does not have Landau pole singularity
 and we can define new  invariant charge as a real part of the originally defined  invariant charge at $p^2 =\exp(i\phi)|p^2|$.
 In one-loop approximation
 new invariant charge has extremum equal to   $ \mp\frac{1}{2\beta_2\phi}$   at $\ln(\frac{|p^2|}{\Lambda^2}) = \pm \phi$.  
 Probably the most natural way is  the definition of  new invariant charge as real part of old invariant charge at negative $p^2$.

\section{The invariant charge in QED with N identical fermions}

In QED with N identical fermions the solution of the renormalization group equation (\ref{QEDb1})
can be represented in the form
\begin{equation}
  \bar{\alpha}(x, \alpha,  N) = \sum_{k=0}^{\infty} (1/N)^k \bar{\alpha}_k(x, \alpha) \,.
\label{QEDb4}
\end{equation}
In the leading order by  $(1/N)$ only bubble type diagrams depicted in
Figure.1  give the contribution  to the invariant charge
equal to
\begin{equation}
  \bar{\alpha}(\frac{p^2}{\mu^2}, \alpha, N) = \frac{-1}{\beta_2\ln(\frac{p^2}{\Lambda^2})} \,,
\label{QEDb5}
\end{equation}
where $ \Lambda^2 = \mu^2\exp(\frac{1}{\beta_2}{\alpha})$.

\begin{figure}[tbh!]
\begin{center}
  \includegraphics[width=.95\textwidth]{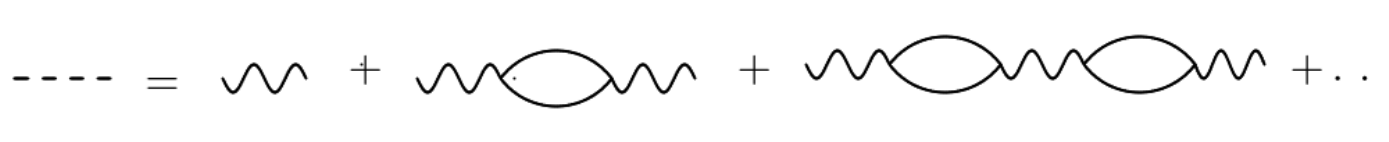}%
\vspace{-3mm}
\caption{The Feynman diagrams    contributing to the ivariant charge in the leading order of $1/N$.}
\end{center}
\label{fig1}
\vspace{-5mm}
\end{figure}

 The $1/N$ corrections to  the
 invariant charge are described by the diagrams depicted in Figure 2.

 \begin{figure}[tbh!]
\begin{center}
  \includegraphics[width=.95\textwidth]{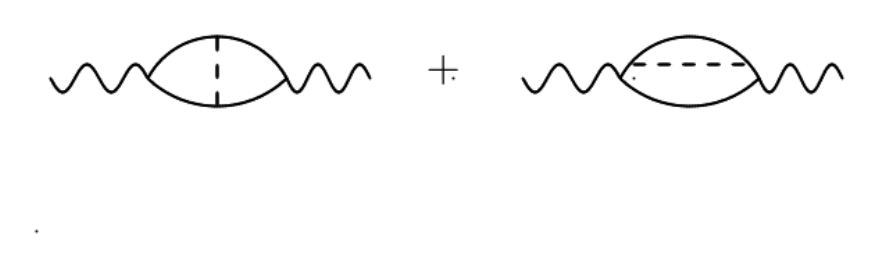}%
\vspace{-3mm}
\caption{The Feynman diagrams    contributing to the
$1/N$ correction for the photon propagator
}
\end{center}
\label{fig2}
\vspace{-5mm}
 \end{figure}
 
   It is useful to define
   the so called Adler $D$-function as
   \begin{equation}
     D(x,\alpha, N) = -x \frac{d}{dx}\Pi(x,\alpha, N) = \frac{\psi(\bar{\alpha}, N)}{\bar{\alpha}^2} \,.
     \label{QEDc}
   \end{equation}
   The $D$-function (\ref{QEDc}) has zero anomalous dimension, i.e. it
   obeys the equation $  [\mu^2 \frac{d}{d\mu^2}+  \psi({\bar{\alpha}}, N)\frac{d}{d]\alpha} ]    D(x,\alpha, N) =  0 $.
     One can find that in QED with the Lagrangian  (\ref{Nqed1})
     \begin{equation}
       D(x = 1, \alpha, N) = \frac{\psi(\alpha, N)}{\alpha^2} = \beta_2 + \frac{\alpha \beta_3}{N} + \sum_{k=2}^{\infty}c_k(N) \alpha^k \,,
   \label{QEDd}
     \end{equation}
     where $\beta_2 = \frac{1}{3\pi}$ and $\beta_3 = \frac{1}{4\pi^2}$  \cite{bog1}.
        For the $D$-function the   $1/N$ expansion is represented in  the form
     \begin{equation}
       D(x, \alpha, N) = 
       \sum^{\infty}_{k =0}(\frac{1}{N})^kD_k(\alpha, x) \,,
\label{QEDm}
     \end{equation}
     where $ D_k(x, \alpha) = \sum_{n =k}^{\infty} \alpha^n \bar{C}_{n,k}(x) $ and $ D_0(x, \alpha) = \beta_2$.

     For the invariant charge $\bar{\alpha}(x, \alpha, N)$  the $1/N$   perturbation theory   coincides with the standard perturbation theory 
     except the use of the effective photon propagator
\begin{equation}
  \frac{\bar{\alpha} (\frac{p^2}{\mu^2}, \alpha)}{p^2} =    -\frac{1}{\beta_2  p^2 \ln(\frac{p^2}{\Lambda^2})} \,
  \label{QEDg}
\end{equation}
 instead of free photon  propagator $\frac{\alpha}{p^2}$.
      For instance,     the $1/N$ correction to  the invariant charge (\ref{QEDa})  is determined by the diagrams
      depicted in Fig.2.
The $1/N$ correction to the photon propagator 
       is ultraviolet divergent as
       $\ln(\ln(M^2))$ while in  standard perturbation theory two-loop  correction  diverges as $\ln(M^2)$. Here $M$ is an ultraviolet cutoff.
      The effective propagator (\ref{QEDg}) decreases in ultraviolet region $p^2 \rightarrow \infty$ faster than local photon
 propagator. As a consequence we find  that the $(1/N)^k$ corrections   to the
 invariant charge without fermion bubbles inside of the diagrams  are ultraviolet finite at $k \geq 2$, see Fig.3.
 In  QED higher order contributions to the Adler function   (\ref{QEDc})   without bubble type subdiagrams   are
 ultraviolet finite \cite{Johnson1, Johnson2}, see also the appendix A. As a consequence  the $1/N$ corrections to  the Adler functions
 without bubble type diagrams are  also ultraviolet finite.

 The diagrams with $1/N$ contribution inside of the fermion loop contain  the ultraviolet divergent $1/N$ subdiagram,
 see Fig.3..   The diagrams  without bubble type subdiagrams
 see Fig.4, are ultraviolet finite.

 \begin{figure}[tbh!]
\begin{center}
  \includegraphics[width=.95\textwidth]{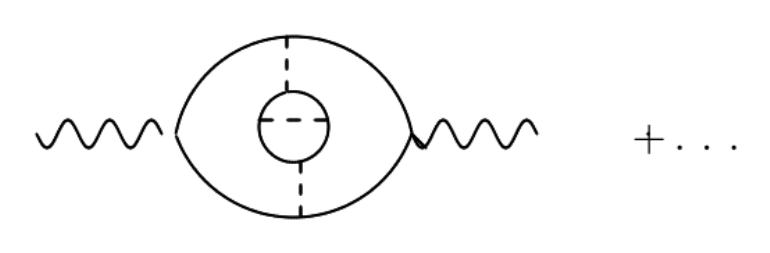}%
\vspace{-3mm}
\caption{The bubble type $1/N$ subdiagram giving infinite contribution  to the
invariant charge}
\end{center}
\label{fig3}
\vspace{-5mm}
\end{figure}

       \begin{figure}[tbh!]
\begin{center}
  \includegraphics[width=.95\textwidth]{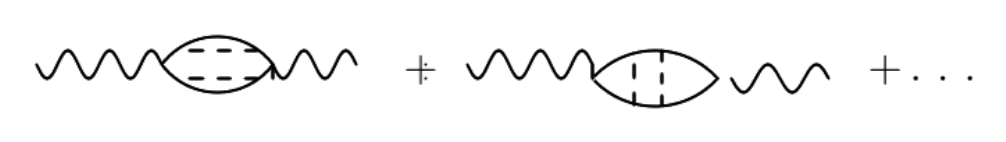}%
\vspace{-3mm}
\caption{Ultraviolet finite $1/N^2$  corrections  to the photon propagator  }
\end{center}
\label{fig4}
\vspace{-5mm}
\end{figure}

   
The knowledge of the D-function (\ref{QEDc}) allows to determine the ultraviolet behaviour of the invariant charge. 
In particular, the $(1/N)^k$ correction to the invariant charge in the ultraviolet region  has the form
\begin{equation}
  \bar{\alpha}_1(\frac{p^2}{\mu^2}, \alpha) = 
 - \frac{\beta_3}{\beta_2^3}
  \frac{\ln(-\ln( \frac{p^2}{\Lambda^2})) } {\ln^2(\frac{p^2}{\Lambda^2})} + O(\frac{1}{\ln^2(\frac{p^2}{\Lambda^2})}) \,.
  \label{QEDb6}
\end{equation}
for $k = 1$ and
\begin{equation}
  \bar{\alpha}_k(\frac{p^2}{\mu^2}, \alpha) \sim  \frac{1}{\ln^{k+1}(\frac{p^2}{\Lambda^2})} \,
  \label{QEDb6a}
\end{equation}
for $k \geq 2$.  The ultraviolet asymptotics of  the $1/N$ correction  (\ref{QEDb6})    to the
invariant charge  is determined by two-loop contribution to the GLM function.

      As it has been mentioned in appendix B  the $1/N$ perturbation theory coincides for both standard QED and the exotic QED
      with   $\alpha_f < 0$ except the use of $\Lambda^2_f$ instead of $\Lambda^2$ in formulae for the effective
       propagator (\ref{QEDg}). It means that the $(1/N)^k$ corrections for both exotic and standard   QED
       coincide except the replacement $\Lambda_f \rightarrow \Lambda$.
       To  get rid of the problems with the Landau pole for  the effective propagator   (\ref{QEDg})
       we 
       consider the effective propagator at complex $\Lambda^2$.
       So we find that in QED with $N$ identical fermions at $N \gg 1$ the ultraviolet asymptotics of the photon propagator
       (invariant charge) is determined by one-loop approximation for photon propagator and the
       leading log approximation (\ref{QEDb5}) is valid.
Another way to understand the results is the following. The photon  wave function renormalization $Z(\infty, \alpha)$ could be
       finite\footnote{The case $Z(\infty, \alpha) \neq 0 $ corresponds to the ultraviolet asymptotics of
         the  photon propagator $D(p^2) \sim O(\frac{1}{p^2})$ while the case of infinite wave function
         renormalization $Z(\infty, \alpha) = 0 $     corresponds
         to $p^2 D(p^2) \rightarrow \infty $ at $p^2 \rightarrow \infty $.  } or infinite. There are arguments \cite{EVIDENCE} that the case of finite wave
       function    renormalization      is not realized. Therefore the single possibility is 
       $Z(\infty, \alpha)  = \pm \infty $. In this case the photon propagator decreases faster than $\frac{1}{p^2}$ at
       $p^2 \rightarrow \infty$.  It  means that  the invariant charge 
       $ \bar{\alpha}(\frac{p^2}{\mu^2}, \alpha) \rightarrow 0 $ at $ p^2 \rightarrow \infty $. The smallness of the invariant charge
       in the ultraviolet region means
       that we can apply the perturbation theory. For positive initial values $\alpha > 0$ the
       GLM function $\psi(\alpha) = \frac{\alpha^2}{3\pi} + O(\alpha^3) $ is positive and the invariant charge is
       increasing function of  $p^2$. So we obtain the contradiction. The single way to get rid of this contradiction is the assumption
       that at some point $p^2 = \Lambda^2 $ the invariant charge  changes the sign and
         $\bar{\alpha}(\frac{p^2}{\mu^2}, \alpha) < 0  $ at $p^2 > \Lambda^2$. Such behaviour takes place  for the invariant charge in
       leading log approximation.   

       \subsection{Modified  $1/N$ perturbation theory}

       The  $1/N$ correction  to  the photon propagator  is described by  
       two-loop diagrams, see Fig.2,  with
       the effective photon  propagator $D_{eff}(p^2, \Lambda^2) =   \frac{1}{-p^2 \beta_{2}  \ln(\frac{p^2}{\Lambda^2})} $ inside of the fermion loop.
           One can find that three-loop diagrams, see Fig.4, and higher loop diagrams
       with the  effective photon  propagators inside the single fermion loop   are ultraviolet finite.
       To obtain ultraviolet finite perturbation theory 
       we  added to  the effective propagator
   (\ref{QEDg})
       ultraviolet  divergent
        $1/N$ corrections, see Fig.2. 
       The improved  effective propagator   $ D^{imp}_{eff}(p^2, \Lambda^2 )$    can be  represented in the form
\begin{equation}
          D^{imp}_{eff}(p^2, \Lambda^2) = \frac{1}{p^2( - \beta_2 \ln(\frac{p^2}{\Lambda^2})  + \frac{1}{N} \Pi_2(\frac{p^2}{\Lambda^2}))} \,.
         \label{imp1a}
\end{equation}
The first terrm in (\ref{imp1a}) is the leading contribution of the diagrams of Fig.1 and the $\Pi_2$ is
the $1/N$ contribution of the diagrams depicted at Fig.2.
One can find that the improved effective propagator   (\ref{imp1a}) leads to ultraviolet finite perturbation theory.
The ultraviolet asymptotics of the improved effective propagator is
\begin{equation}
   D^{imp}_{eff}(p^2, \Lambda^2) =  \frac{1}{p^2( - \beta_2 \ln(\frac{p^2}{\Lambda^2})  +
    \frac{\beta_3}{N\beta_2} \ln(-\ln(\frac{p^2}{e\Lambda^2}))   )} \,,
   \label{photas}
\end{equation}
where $e = 2.71928$.
For the improved effective propagator (\ref{photas}) we used the normalization condition
 $ (D^{imp}_{eff}(\frac{\Lambda^2}{\mu^2}, \alpha)\Lambda^2)^{-1} = 0  $.

\section{Conclusions}

In this paper we
investigated ultraviolet behavior of the invariant charge
in QED with massless fermions. In QED the invariant charge is proportional to the transverse part of the photon propagator
$\bar{\alpha}(\frac{p^2}{\mu^2}, \alpha, N) =
\alpha p^2 D_{tr}(p^2, \mu^2, \alpha, N)$.
   As it is well known in the leading log approximation photon propagator has Landau pole
singularity. We considered photon propagator at complex  $p^2 $. For complex $p^2 $ including negative $p^2$
photon propagator does not have Landau pole singularity. For QED with N identical fermions
the ultraviolet behavior of $(1/N)^k$ correction to the photon propagator  coincides with
ultraviolet behaviour of the photon propagator  in  nonphysical but asymptotically free QED with imaginary charge. We  proposed modified ultraviolet finite $1/N$ perturbation theory for
photon propagator. Our results give hint in favour of the hypothesis that in other non asymptotically free models like scalar QED or
Wess-Zumino model the ultraviolet behavior of the invariant charge coincides with the leading log approximstion.

\section{Acknowledgments}
I  am  indebted to  prof. A.L. Kataev for useful conversations and for pointing out to me some references. Also I thank 
  collaborators of the INR theoretical department   
for discussions and critical comments.

\subsection{Appendix A. The expansion to a series in  N in QED}

  Consider QED with N identical fermions and  with the replacement $ \frac{e}{\sqrt{N}}
  \rightarrow e $   in  the Lagrangian (\ref{NQED1}).
  Instead of the $1/N$ expansion in this appendix  we  consider the expansion to a series in   the $N$ parameter. In
realistic models the paramete $N$ is an integer non negative number.  However we can analytically extend all formulae to
the case of non integer $N$ as in the case of dimensional regularization.
The expressions for the invariant charge and the GLM functions  can be  represented in the form
\begin{equation}
  \bar{\alpha}(x, \alpha, N) = \alpha + \sum_{k =1}^{\infty} N^k \bar{\alpha}^{'}_{k }(x,\alpha) \,,
    \label{NQED1}
  \end{equation}
   \begin{equation}
     \psi(\alpha, N) = \sum_{k=1}^{\infty}N^k\psi^{'}_{k}(\alpha) \,,
     \label{NQED2}
   \end{equation}
   where
     $ \bar{\alpha}^{'}_{k}(x=1,\alpha) = 0$.
   Feynman diagrams with N fermion bubbles give the contribution to
   $ \bar{\alpha}^{'}_{k}(x,\alpha)    $ and $\psi^{'}_{k}(\alpha)    $. As a consequence of the renormalization group equation
   \begin{equation}
     x \frac{d\bar{\alpha}(x, \alpha, N)}{dx} = \psi(\bar{\alpha},N) \,,
       \label{NQED3}
       \end{equation}
   the $N$ expansions of the invariant charge (\ref{NQED1}) and the GLM function (\ref{NQED2}) one can find that
   \begin{equation}
     \bar{\alpha}^{'}_{k}(x,\alpha) = \sum_{l=1}^{l = k }c_l(\alpha)(\ln(x))^l \,.
     \label{NQED4}
   \end{equation}
   In particular, we find that
 the  Feynman diagrams for photon propagator  with the single fermion bubble are
   logarithmically divergent and 
      \begin{equation}
 \bar{\alpha}^{'}_{k=1}(x,\alpha) = c_1(\alpha)\ln(x) \,.
     \label{NQED5}
      \end{equation}
Moreover the contrubution to the D-function is ultraviolet finite, namely
      $D(x, \alpha, N) = N \frac{c_1(\alpha)}{\alpha^2} + O(N^2) $.

\subsection{Appendix B. QED with imaginary charge}
     Consider  QED with imaginary charge $e^{'} = ie $. Such model is nonphysical since the interaction  Lagrangian
     is nonhermitean and the property of the unitarity is lost\footnote{The one-loop approximation (\ref{C2}) for the invariant charge
       is negative at $p^2 > \Lambda^2 $ }. Nevertheless we can use the perturbation theory as in standard case.
     For the model with
     the imaginary charge the   coupling constant $\alpha^{'} = \frac{(e^{'})^2}{4\pi} < 0 $ is negative. As a consequence
     we have the asymptotic freedom for negative  coupling constant $ \alpha' $. The GLM function
     for the model with imaginary charge
     coincides for the GLM function $\psi(\alpha, N) $ at negative $\alpha $. 
        It means that for such exotic model the perturbation theory is reliable in the ultraviolet region $ p^2 \gg \Lambda^2_f $ where the invariant charge 
          $\bar{\alpha}(x, \alpha') = 
     \frac{1}{-\beta_2\ln(\frac{p^2}{\Lambda_f^2})}$ and  $\Lambda^2_f = \mu^2 \exp(\frac{3\pi}{\alpha'})$.
       Due to  the asymptotic freedom of the QED with imaginary charge we  can reliably calculate the ultraviolet behaviour
     of the $(\frac{1}{N})^k$ correction to the invariant charge.
      The use of the  renormalization group leads to  the ultraviolet asymptotics
     \begin{equation}
     D_k(\frac{p^2}{\mu^2},  \alpha^{'})   \sim (\ln(\frac{p^2}{\Lambda^2_f}))^{-k} \,
       \label{QEDl}
     \end{equation}
     for the $(1/N)^k$ correction to the $D$-function. Due to
       the $D$-function  definition (\ref{QEDc})   the ultraviolet behaviour of the $(1/N)^k$ correction to
     the invariant charge is determined by the formulae
     \begin{equation}
       \bar{\alpha}_k(\frac{p^2}{\nu^2}, \alpha^{'})   \sim (\ln(\frac{p^2}{\Lambda^2_f}))^{-k  - 1} \,
       \label{QEDm1}
       \end{equation}
       at $k  > 1 $ and
      \begin{equation}
       \bar{\alpha}_1(\frac{p^2}{\mu^2}, \alpha)   \sim   \frac{\ln (\ln(\frac{p^2}{\Lambda^2_f}))}{ (\ln(\frac{p^2}{\Lambda^2_f}))^2      } \,
       \label{QEDm2}
       \end{equation}
      at $k = 1$. For the invariant charge as a whole we find the following ultraviolet asymptotics:
      \begin{equation}
        \bar{\alpha}(\frac{p^2}{\Lambda^2_f}, N)   =
        \frac{1}{(-\beta_2 \ln(\frac{p^2}{\Lambda^2_f})  + \frac{\beta_3}{N}
         \ln(e \ln(\frac{p^2}{\Lambda^2_f})) + o(1)  )} \,. 
        \label{QEDm11}
      \end{equation}
      For the ultraviolet asymptotics (\ref{QEDm11}) we used the normalization condition
      $\bar{\alpha}^{-1}(\frac{p^2}{\Lambda_f}  = 1, N) = 0$.
   
      Our key observation is that the $1/N$ perturbation theory coincides for both standard QED and the exotic QED
      with   $\alpha^{'} < 0$ except the use of $\Lambda^2_f$ instead of $\Lambda^2$ in formulae for the effective
       propagator (\ref{QEDg}). It means that the $(1/N)^k$ corrections for both exotic and standard   QED
       coincide except the replacement $\Lambda_f \rightarrow \Lambda$.
       Note that to  get rid of the problems with the Landau pole for  the effective propagator   (\ref{QEDg})
       we 
       consider the effective propagator at complex $\Lambda^2_f$.


\newpage
\begin{thebibliography}{99}
\bibitem{EVIDENCE} D.E.J.Callaway, Phys.Rept. {\bf 167}, 241 (988).
\bibitem{bog1} N.N.Bogoliubov and D.V.Shirkov, Introduction to the theory of quantized fields(Interscience, New York, 1959) Chs. VIII, IX.
\bibitem{N1} H.E.Stanley, Phys.Rev. {\bf 176}, 718 (1968).
\bibitem{N2} K.Wilson, Phys.Rev. {\bf D7}, 2911 (1973).
\bibitem{N3} M.Moshe and J.Zean-Justin, Physics Reports {\bf v.385}, 69 (2003).
\bibitem{KL1} G.Kallen, Helv.Phys.Acta {\bf 25} 417 (1952). 
\bibitem{KL2}   H.Lehmann, Nuovo Cim. {\bf 11}, 342 (1954).
\bibitem{Berestetsky} E.M.Liftshits and L.P.Pitaevsky, Relativistic quantum field theory, vol.2, Nauka 1971, Moscow (in russian).
\bibitem{Gel} M.Gell-Mann and F.Low, Phys.Rev. {\bf 95}, 1300 (1954).
\bibitem{Pet} E.C.Stueckelberf and A.Peterman, Helv.Phys.Acta {\bf 26}, 499 (1953).
\bibitem{Kr1} N.V.Krasnikov, Phys.Lett. {\bf 105B}, 212 (1981).
\bibitem{Kr2} N.V.Krasnikov, Nucl.Phys. {\bf B192}, 497 (1981).
\bibitem{Yam} Y.Yamagishi, Phys.Rev. {\bf D25}, 464 (1982).
\bibitem{Kr3} N.V.Krasnikov, arXiv:2510.14563(2025).
\bibitem{QEDN1} R.Coquereaux, Phys.Rev. {\bf D23}, 2776 (1981).
\bibitem{QEDN2} D.Espriu, A.Palanques-Mestre, P.Pascual and R.Tarrach, \\
  Z.Phys. {\bf C13}, 153 (1982).
\bibitem{QEDN3} A.Palanques-Mestre and P.Pascual, Commun.Math.Phys. {\bf 95}, 277 (1984).
\bibitem{QEDN4} D.J.Broadhurst, Z.Phys. {\bf C58}, 339 (1993).
\bibitem{LINDE} D.Kirsnitz and A.Linde, Phys.Lett. {\bf B73}, 323 (1978).  
\bibitem{Kat1} S.G.Gorishny, A.L.Kataev, S.A.Larin and L.R.Surguladze, Phys.Lett. {\bf B256}, 81 (1991).
\bibitem{Kat2} P.A.Baikov, K.G.Chetyrkin, J.H.Kuhn and J.Rittenger, JHEP {\bf 07}, 017 (2012).
\bibitem{Kat3} A.L.Kataev and S.A.Larin, Pisma v ZhETF  {\bf v.96}, No 1, 64 (2012);\\
  arXiv:1205.2810.
\bibitem{PIVO} N.V.Krasnikov and A.A.Pivovarov, Phys.Lett. {\bf B116}, 168 (1982).
\bibitem{PIVO1} N.V.Krasnikov and A.A.Pivovarov, Mod.Phys.Lett. {\bf A11}, 835 (1996).
 \bibitem{Johnson1} K.Johnson, R.Willey and M.Baker, Phys.Rev. {\bf 163}, 1699 (1967).
\bibitem{Johnson2} M.Baker and K.Johnson. Phys.Rev. {\bf 183} 1292 (1969).


\end{thebibliography}
\end{document}